\documentclass[fleqn,twoside]{article}
\usepackage{latexsym}
\usepackage{amssymb}
\usepackage{espcrc2}

\newcommand{\rf}[1]{(\ref{#1})}
\newcommand{\bea}{\begin{eqnarray}}
\newcommand{\eea}{\end{eqnarray}}

\renewcommand{\d}{\mbox{d}}
\newcommand{\g}{\gamma}

\newcommand{\sg}{\sigma}

\newcommand{\ra}{\rangle}
\newcommand{\la}{\langle}

\usepackage{graphicx}
\usepackage[figuresright]{rotating}
\usepackage{epsfig}

\newcommand{\AmS}{{\protect\the\textfont2
  A\kern-.1667em\lower.5ex\hbox{M}\kern-.125emS}}
\newcommand{\beq}{\begin{equation}}
\newcommand{\eeq}{\end{equation}}
\newcommand {\beqa}{\begin{eqnarray}}
\newcommand {\eeqa}{\end{eqnarray}}

\title{Real--time dynamics of a hot Yang-Mills theory:
a numerical analysis}

\author{J. Ambj\o rn\address[NBI]{The Niels Bohr Institute, Blegdamsvej
        17, DK-2100 Copenhagen \O, Denmark},
        K. N. Anagnostopoulos \address{Department of Physics,
        University of Crete, GR-710 03 Heraklion, Greece}
        and
        A. Krasnitz \address{CENTRA and Faculdade de Ci\^encias e
        Tecnologia, Universidade do Algarve, Campus de Gambelas,
        P--8000, Faro, Portugal} 
        }
       
\begin{document}

\begin{abstract}
We discuss recent results obtained from simulations of high
temperature, classical, real time dynamics of $SU(2)$ Yang-Mills theory at
temperatures of the order of the electroweak scale. Measurements of gauge 
covariant and gauge
invariant autocorrelations of the fields indicate that the
ASY-B\"odecker scenario is irrelevant at these temperatures.
\end{abstract}

\maketitle

High temperature baryon number non conservation in the electroweak
theory caused by the anomaly of the baryonic current is important for
the understanding of the origin of baryonic matter in the universe.
After the work of Ref.\ \cite{krs} it was realized that the baryon
number non-conservation, caused by thermal field fluctuations between
gauge-equivalent vacua with different winding numbers could be large
in the unbroken phase of the electroweak theory. This can be tested by
performing real time simulations of hot gauge theories, a task we
still do not know how to perform from first principles. It is
possible, however, to use {\it classical} thermal gauge theory in
order to address the question \cite{grs}. The result is that
transitions between vacua with different winding numbers is
unsuppressed at high temperatures in the unbroken phase of the
electroweak theory \cite{aaps}. The effect of correctly incorporating
ultraviolet thermal fluctuations has been addressed in the the theory
of hard thermal loops and the effective small-momentum, low frequency
theory of B\"odeker (ASY-B) \cite{htl}. 
The soft classical 
fields (momentum $k \leq gT$) couple to hard currents 
according to 
\begin{equation}
\label{1.1}
\dot{\mathbf{E}}= {\mathbf D}\times{\mathbf B}- {\mathbf J}_{\rm hard},
\end{equation}
where ${\mathbf J}_{\rm hard}= \sg {\mathbf E} + \mbox{\boldmath
$\xi$}$, and where the effective noise term {\boldmath $\xi$} is
determined by the fluctuation-dissipation theorem 
$\la \xi\xi\ra=2T\sg$.
In
\rf{1.1} $\sg$ denotes the so-called color conductivity, which to a
leading $\log$ approximation is given by $\sg = \frac{m^2}{\g}$, 
where $m$ denotes the Debye screening mass and $\g$ the hard gauge fields
damping rate.  
High
frequency magnetic modes couple to low frequency ones producing a new
time scale for the magnetic fluctuations. Although their length scale
is of order $1/g^2 T$ (non-perturbative) their lifetime is of order
$1/g^4T\ln(g^{-1})$. 
This picture has been verified in real time
computer simulations \cite{num} where hard currents have been
implemented in various ways, by measuring primarily the sphaleron
rate. The latter is believed to be dominated by classical thermal
fluctuations and one can hope that simulations of the classical
theory, possibly corrected by ASY-B effective theory, can determine it
correctly. It is, however, possible that at the electroweak scale
($\alpha\approx 1/30$, $g\approx 0.65$) and at temperatures close to
the electroweak phase transition the ASY-B theory cannot be
applied. This can be checked in computer simulations by measuring real
time gauge correlators and color conductivity and check whether their
long wavelength, low frequency part shows the behavior predicted by
ASY-B, this way making direct contact with perturbation theory. If we
do not obtain agreement, one could be tempted to conclude that the
theory is valid at even higher temperatures and that one should {\it
not} try to match the sphaleron rate to formulas based on the validity
of the ASY-B theory. Precisely because the sphaleron rate is a
non-perturbative quantity, there is no easy way to disentangle the
perturbative ASY-B damping from genuine non--perturbative effects. We
also note that topology on the lattice is ill-defined and requires
special treatment.  In Ref.\ \cite{AAK} we attempted to address this
issue by precisely measuring objects like color conductivity and field
autocorrelators. Our finding is that we see no sign of the ASY-B
scenario in high-temperature, classical $SU(2)$ theory in a regime
roughly corresponding to the electroweak scale. We stress that our
results are not in contradiction with the results by Moore and
Rummukainen \cite{num} which give zero continuum sphaleron rate, but
do not rule out a finite classical rate in the continuum.
\begin{figure}[hbt]
 {\epsfig{file=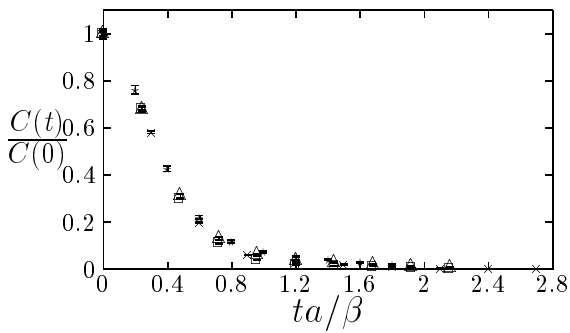,width=75mm}}
\caption{The autocorrelator
$\langle {\bf D}\times{\bf B}({\bf x},t)\cdot{\bf D}\times{\bf B}({\bf x},0)
\rangle$ versus time
$t$ in units of $(g^2T)^{-1}$ for $\beta=8.33$ (crosses), 
$\beta=10.0$ (squares), $\beta=12.5$ (triangles), and $\beta=15.0$ (stars).}
\label{fig1}
\vspace{-6mm}
\end{figure}
\begin{figure}[hbt]
 {\epsfig{file=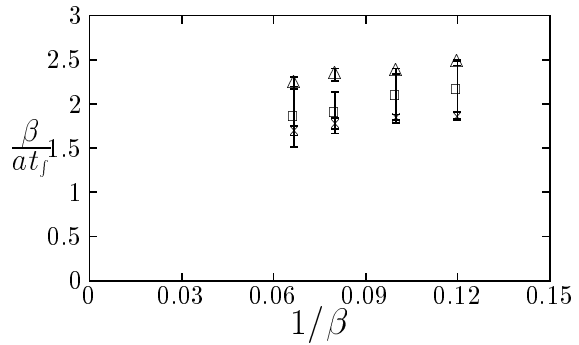,width=75mm}}
\caption{Inverse integral autocorrelation time in 
units of $g^2T$ plotted against
$g^2Ta$ for
$\langle {\bf D}\times{\bf B}({\bf x},t)\cdot{\bf D}\times{\bf B}({\bf x},0)
\rangle$ (triangles), $\langle {\bf B}({\bf x},t\cdot{\bf B}({\bf x},0)\rangle$
(crosses), and
$\langle B_i^2({\bf x},t)B_i^2({\bf x},0)\rangle-
\langle B_i^2({\bf x},t)\rangle\langle B_i^2({\bf x},0)\rangle$ (squares).}
\label{fig4}
\vspace{-6mm}
\end{figure}
\begin{figure}[hbt]
 {\epsfig{file=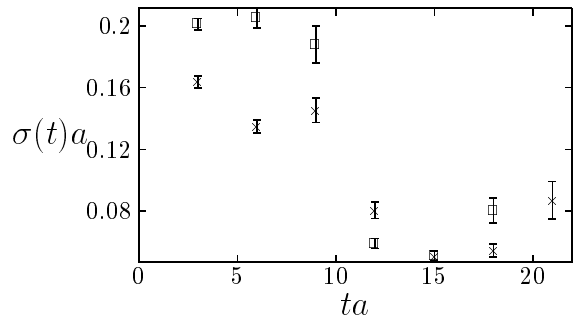,width=75mm}}
\caption{The color conductivity in lattice units 
versus time
$t$ in lattice units for $\beta=10$ (crosses) and
$\beta=12.0$ (squares).}
\label{fig6}
\vspace{-6mm}
\end{figure}

Simulations of the classical theory are performed by generating
thermal field configurations for pure $SU(2)$ Yang-Mills theory on a
lattice for given temperature $T$ and then lettting the system evolve
according to the classical equations of motion \cite{K}. We worked in
the temporal gauge where the electric field ${\mathbf E}({\mathbf
x},t)$ is the conjugate momentum to ${\mathbf A}({\mathbf
x},t)$. Since the cutoff of the classical theory is the lattice
spacing whereas the thermal fluctuations of the full quantum
theory are cut off at scale $T$, we need to filter out the large
momentum components of the fields by cooling \cite{cool}, which leads
to an exponential decay of high-frequency modes. In our simulations we
used cooling times $\tau$ such that $(g^2T)^2\tau=3.84$ (also $2.56$
for $\beta=10$). Variance of $\tau$ had very small effect on real time
behavior of correlators.  The lattice temperature $\beta\equiv 4/(g^2
T a)$ was chosen within a range such that the perturbative Debye mass
$m_D$ to $g^2T$ of the classical theory is close to that of the full
$SU(2)$ Yang-Mills theory at electroweak temperatures $T\sim
100$GeV. Since we are interested in the dynamics of fields with
momenta of the order of $g^2T$, the dimensionless combination
$L/(\beta a)$ should be large enough in order to avoid finite-size
effects. Most of our simulations were performed at $L/(\beta
a)=2.4$. We verified that variations of $L/(\beta a)$ around that
value did not have a measurable effect.
We measured unequal time correlators of the fields of the form
$C_{12}(t) \equiv \frac{1}{V} \int \d^3 x \;
\la {\cal O}_1({\mathbf x},t) {\cal O}_2({\mathbf x},0)\ra$
where the ${\cal O}_i({\mathbf x},t)$ are ${\mathbf D}\times {\mathbf
B}({\mathbf x},t)$ or ${\mathbf B}({\mathbf x},t)$. In particular we
measured 
\begin{equation}
\label{colcon}
\sg(t) \equiv\frac{\int \d^3 x\;\langle {\bf D}\times{\bf B}({\bf
x},0)\cdot{\bf D}\times{\bf B}({\bf x},t)\rangle}{\int \d^3 x
\;\langle {\bf D}\times{\bf B}({\bf x},0)\cdot {\bf E}({\bf
x},t)\rangle}\, ,
\end{equation}
where $\sg(t)\to \sg$ for large enough $t$ and $\sg$ is the color
conductivity. We also studied plasmon excitations by measuring the
color charge density $\rho\equiv\mathbf{D\cdot E}$.  
An equivalent gauge-invariant definition for correlators of these adjoint 
objects requires introducing a straight Wilson 
line connecting $({\bf x},0)$ to $({\bf x},t)$. This Wilson line becomes
an identity in the temporal gauge.
In order to test the effect of this Wilson line
on the characteristic time scale of the correlators we also study a
truly gauge-invariant object, $B^2({\bf x},t)$, and we find no
difference with gauge covariant quantities.
Our definition of the color conductivity is similar to that of
Arnold and Yaffe \cite{htl}, which is also based on the effective theory
in the temporal gauge.

In Fig.\ \ref{fig1} we show autocorrelators for the
$B$-field, with time measured in units of $(g^2T)^{-1}$. We note that the 
curves corresponding to different values
of $\beta$ coincide as long as the correlators retain a substantial
portion of their original value. We also introduce the integral
autocorrelation time defined for an autocorrelator $C(t)$ as
$t_{\int}\equiv (C(0))^{-1}\left(\int_0^\infty C(t){\rm
d}t\right)$. In Fig.\ \ref{fig4} we plot the dimensionless quantity
$4/(g^2Tt_{\int})$ as a function of $1/\beta=g^2Ta/4$.  Remarkably, in
all three cases $t_{\int}$ turns out to be of the order of $g^2T$ and
shows little dependence on the lattice spacing throughout the range
considered. There is therefore no evidence that in this range of the
lattice spacings our cooled autocorrelators follow the ASY-B scenario,
wherein the expected behavior is $t_{\int}\propto 1/(g^4T^2a)$, up to
logarithmic corrections. In the case of the color charge
autocorrelator $\langle {\bf D}\cdot{\bf E}({\bf x},t){\bf D}\cdot{\bf
E}({\bf x},0)\rangle$ the time scale for the color charge correlation
is proportional to the lattice spacing and does not depend on $g^2T$.
This result can be contrasted with perturbative predictions.  One
would expect that the color-charge autocorrelator is dominated by the
plasmon mode, whose frequency in the classical theory is of the order
$g\sqrt{T/a}$ and whose decay rate is of the order $g^2T$. We observe
none of these properties in the range of lattice spacings considered.
Finally, we attempted to determine color conductivity $\sigma$. As
Figure \ref{fig6} 
demonstrates, this attempt failed in two ways. First of all,
$\sigma(t)$ does not appear to approach a constant for times in excess
of the expected autocorrelation time of the noise term ${\bf \xi}$
(and far in excess of the measured autocorrelation time of the
noise). Secondly, the numerical value of $\sigma(t)$ is very small
(less than $0.25/a$) compared to the value expected in the ASY
scenario ($a\sigma\approx 15$). 

\end{document}